\documentclass[twocolumn,pra,superscriptaddress,showpacs,aps,floatfix]{revtex4-1}
\usepackage{color}
\usepackage{amsmath}
\usepackage{amssymb}
\usepackage{txfonts}
\usepackage{graphicx}
\usepackage{epstopdf}
\usepackage[unicode]{hyperref}
\hypersetup{pdftitle={2D SOC BEC},
 pdfauthor={},
 pdfsubject={2D SOC BEC},
  colorlinks=true,
    linkcolor=red,
    citecolor=magenta,
    filecolor=black,
    urlcolor=blue}


\begin{document}

\title{An Improved Standard Model Comes with Explicit CPV and Productive of BAU}

\author{Chilong Lin}
\email{lingo@mail.nmns.edu.tw}

\affiliation{National Museum of Natural Science, 1st, Guan Chien RD., Taichung, 40453 Taiwan, ROC}

\date{Version of \today. }

\begin{abstract}
In this manuscript, we present an explicit way to describe the violation of CP symmetry in the standard model (SM) of electroweak interactions.
In such a way, complex Cabbibo-Kobayashi-Maskawa (CKM) matrices are achieved which stand for the violation of CP symmetry.
At the beginning, two necessary but not sufficient conditions for yielding a complex CKM matrix are stated as criteria.
Then we found an interesting condition between the real and imaginary parts of a Hermitian $3\times 3$ matrix may provide extra relations among its parameters and reduce the number of them from eighteen down to five.
In previous investigations, this can be done only down to nine.
With another assumption among some of those parameters, the mass-matrix pattern is further simplified so as to be diagonalized analytically and in consequence four matrices which reveal $S_N$ symmetries among or between quark generations are obtained.
In some of such $S_N$-symmetric cases, the derived CKM matrices are complex which indicate that CP symmetry is violated accordingly.
Taking the Jarlskog invariant as an estimate of the CPV strength, the value predicted by this model is orders stronger than the empirical value detected experimentally.
However, that happens to fill partly the gap between the cosmologically observed amount of Baryon Asymmetry of the Universe (BAU) and that current Standard Model of particle physics predicts. It also proves the long suspected existence of BAU-productive eras in early universe if some fermions were indistinguishable, i.e., $S_N$-symmetric, under circumstances of extremely high temperatures.

\end{abstract}
\maketitle


\section{Introduction}

Since the first detection of CP violation (CPV) in 1964 \cite{Christenson1964}, the issue of how CP symmetry was violated, attracting our interest very much.
However, for over fifty years, an explicit way to describe the violation of CP symmetry is still obscure. If we analyze the theoretical origin of CPV from its fundamental, considering only the quark sector here, the patterns of quark-mass matrices $M$ are obviously the keys to ignite such a violation.
Thus, we would like to analyze the CPV problem by starting from the most general pattern of an $M$ matrix and then simplify it to a manageable level step by step. \\

In the standard model (SM) of electroweak interactions "Direct" CPV is allowed if a complex phase appears in the Cabbibo-Kobayashi-Maskawa (CKM) matrix describing quark mixings,
or the Pontecorvo-Maki-Nakagawa-Sakata (PMNS) matrix describing lepton mixings.
In SM such a complex phase can only be yielded by ingeniously arranging the Yukawa couplings between fermions and Higgs fields and then diagonalizing the Yukawa-coupling matrix directly. However, after decades such a satisfactory Yukawa-coupling matrix is still obscure.. \\

Theoretically there is another way in SM to extract a complex phase from the vacuum expectation value (VEV) of its only Higgs doublet.
 However, such a phase can always be absorbed into redefinitions of quark fields in the SM.
 Thus, an extension of SM with one extra Higgs doublet was proposed in \cite{TDLee1973} which is usually referred to as the Two-Higgs-doublet model (2HDM).
 In this way, people expect the phases of two VEVs may unlikely be rotated away simultaneously if there were a nontrivial phase difference between them.
 However, such an extra Higgs doublet not only failed to solve the CPV problem, but also brought in an extra problem,
 the flavor-changing neutral current (FCNC) problem at tree level. \\

In fact, most of the derivations in this manuscript were originally proposed for solving the FCNC problem and CPV problem in 2HDMs \cite{Lin1988}.
However, we find they apply to the SM as well. As SM alone is enough to describe the violation of CP symmetry,
why should one bother to deal with the extra Higgs doublet and problems it brings in?
Thus, this manuscript is to be devoted to the theoretical ignition of CPV in the SM alone. \\

The Yukawa couplings of $Q$ quarks in SM are usually written as
\begin{equation}
-{\cal L}_Y ~=  \bar{Q_L} Y^{d} \Phi d_R +  \bar{Q_L} \epsilon Y^{u} \Phi^{\ast} u_R + h. c.,
 \end{equation}
where $Y^{q}$ are $3 \times 3$ Yukawa-coupling matrices for quark types $q=u$, $d$ and $\epsilon$ is the $2 \times 2$ antisymmetric tensor.
$Q_L$ is left-handed quark doublets, and $d_R$ and $u_R$ are right-handed down- and up-type quark singlets in their weak eigenstates, respectively. \\

When the Higgs doublet $\Phi$ acquires a VEV, $\langle \Phi \rangle = \left( \begin{array}{cc} 0 \\ v/\sqrt{2} \end{array}\right)$, Eq.(1) yields mass terms for quarks with $M^q=Y^q v / \sqrt{2}$ the mass matrices for $q$ = $u$, $d$.
The physical states are obtained by diagonalizing $Y^q$ with four unitary matrices $U^q_{L,R}$, as $M^q_{diag.} = U^q_L M^q U^q_R  = U^q_L (Y^q v/\sqrt{2} ) U^q_R $.
As a result, the charged-current $W^{\pm}$ interactions couple to the physical $u_L$ and $d_L$ quarks with couplings given by
\begin{equation}
-{\cal L}_{W^{\pm}} ~= {g \over \sqrt{2}} (\bar{u_L},~\bar{c_L},~\bar{t_L})~\gamma^{\mu} W^+_{\mu}~ V_{CKM} \left( \begin{array}{ccc} d_L \\ s_L \\ b_L \end{array}\right) + h. c.,
 \end{equation}
 where
\begin{eqnarray}
V_{CKM} = U^u_L U^{d \dagger}_L =\left( \begin{array}{ccc} V_{ud} & V_{us} & V_{ub}  \\ V_{cd} & V_{cs} & V_{cb}  \\ V_{td} & V_{ts} & V_{tb} \end{array}\right),
\end{eqnarray}
is the CKM matrix describing quark mixings.
Hereafter, we will neglect the sub-index $L$ in quark fields $q_L$ and the unitary matrices $U_L$ if unnecessary. \\

If one demands a complex $V_{CKM}$, following two conditions are necessary: \\

${\bf Condition~1}$. At least one of $U^u$ or $U^d$ is complex, i.e., $U^u$ and $U^d$ musn't be both real. \\

${\bf Condition~2}$. Even if both them were complex, they must not be the same, i.e., $U^u \neq U^d$.
Otherwise, $V_{CKM}$ in Eq.(3) will be a ${\bf 1}_{3\times 3}$ identity matrix which is obviously real. \\

As $V_{CKM}=U^u_L \cdot U^{d\dagger}_L$ is the product of $U^u$ and $U^{d \dagger}$, obviously these two U matrices decide everything of $V_CKM$ including if it were complex.
As $U^u$ and $U^d$ are unitary matrices which diagonalize mass matrices $M^u$ and $M^d$, respectively.
It is obvious they are objects derived from $M^u$ and $M^d$.
Or, we may say the patterns of $M^u$ and $M^d$ are keys to ignite CPV in the standard model.
Thus, we will start the investigation from a most general pattern of $M$ matrix and then put in constraints to see what will happen to the CPV under various circumstances. \\

In section II, the most general pattern of $M$ matrix is given as a start point of the investigation.
If $M$ were Hermitian, an interesting condition Eq.(6) comes in if its real and complex parts can be diagonalized respectively by a same unitary matrix and the number of independent parameters in the $M$ matrix is thus reduced from eighteen down to five.
While in previous similar researches this can be done only down to nine with the same assumption. \\

Since only three quark masses are now given in a quark type, an $M$ matrix with five unknowns is still too complicated to be diagonalized analytically.
Thus, an assumption $A=A_1=A_2=A_3$, where $A_1$, $A_2$ and $A_3$ are diagonal elements of the most general M matrix to be given in section II,
is employed to further simplify its pattern.
Such an assumption gives us four analytically diagonalizable $M_k$ matrices and correspondingly four complex $U_k$ matrices,
where $k$ = 1 to 4 indicate to which case they correspond.
These multiple $U_k$ matrices enable us to satisfy both necessary conditions mentioned above for yielding a complex CKM matrix. \\

Even we have already a way to describe the violation of CP symmetry within the standard model.
However, as to be demonstrated in section III, the CPV derived in such $S_N$-symmetric cases are orders stronger than the value current SM can provide and orders weaker than the one needed to account for the cosmologically observed Baryon Asymmetry of the Universe (BAU) [and references therein] \cite{Tranberg2010}.
The derived Jarlskog invariant ${\it J} \sim 0.171$, which is usually used to estimate the CPV strength \cite{Jarlskog1985},
is about four orders stronger than the experimentally detected value ${\it J}=3.00_{-0.09}^{+0.15} \times 10^{-5}$ \cite{Zyla2020} which corresponds to the BAU amount of order $\eta \sim 10^{-20}$ \cite{Canetti2012} and at least six orders weaker than the quantity $\eta = {N_B \over N_{\gamma}} \vert_{T=3K} \sim 10^{-10}$ \cite{Zyla2020, Canetti2012, Ade2016} estimated from the baryon/photon ratio observed in present universe.
At first glance such a CPV strength looks like a defect of the model.
However, if we consider such $S_N$ symmetries should exist only under circumstances of extremely high temperatures.
It is not strange that CPV derived in this manuscript is different to the one detected in present experiments.
Besides, it indicates that BAU we see today could be remnant left over in some early $S_N$-symmetric eras of our universe, at least part of them. \\

Conclusions and discussions are to be given in section IV. \\

\section{Analytically Diagonalizable $M$ matrices and their $U$ matrices}

As mentioned in Eq.(3), $V_{CKM}$ is an inner product of two $U$ matrices, one $U^u$ from the up-type quarks and one $U^{d\dagger}$ from the down-type quarks, while they are respectively derived from the mass matrices $M^u$ and $M^d$.
Obviously these two $M$ matrices are key factors to determine if $V_{CKM}$ were complex.
Based on this, the most orthodox way is to start from the most general pattern of $M$ matrices and then diagonalize them to achieve corresponding $U$ matrices. \\

In the SM with three fermion generations and one Higgs doublet, the mass matrix of any specific fermion type must has the general pattern as
\begin{eqnarray}
M =  \left( \begin{array}{ccc} A_1 +i D_1 & B_1+ i C_1 & B_2+ i C_2   \\
 B_4 + i C_4 & A_2 +i D_2 & B_3 + i C_3  \\ B_5 + i C_5 & B_6 + i C_6 & A_3 +i D_3 \end{array}\right) ,
\end{eqnarray}
which contains eighteen parameters and $A$, $B$, $C$ and $D$ parameters are all real.   \\

Theoretically, the most orthodox way to derive the CKM matrix is to diagonalize one such matrix for up-type quarks and one for down-type quarks and then derive their corresponding $U$ matrices.
However, such a pattern is obviously too complicated to be diagonalized analytically. Not to mention the $V_{CKM}$ thus derived contains thirty six parameters in total, eighteen from the up-type quarks and eighteen from the down-type quarks. \\

In general, physicists employ various constraints to simplify the matrix pattern to a manageable level.
For instance, the Fritzsch ansatz (FA) ~\cite{Fritzsch1978, Fritzsch1979} and its subsequent developments like Cheng-Sher ansatz (CSA) \cite{Cheng1987}, Du-Xing ansatz (DXA) \cite{DuXing1993}, combination of the Fritzsch and the Du-Xing ansatz (DFXA and FDXA), combination of different assignments in the Du-Xing ansatz ($\tilde{X} A$), Non-mixing Top quark Ansatz (NTA) [and references therein]\cite{Carcamo2007} and Fukuyama-Nishiura ansatz (FNA) [and references therein]\cite{Matsuda2000} imposed several zeros in the $M$ matrix as $ad~hoc$ constraints to reduce the number of parameters in it.
The goal of all of them is just to simplify the $M$ pattern to a manageable level.
However, constraints so strong are in fact unnatural and unnecessary.
As to be demonstrated in what follows, a very weak assumption that $M$ matrices are Hermitian is already enough to yield a very simple $M$ pattern and it includes almost previous ansatz as special cases in it. \\

It's interesting that action of "different assignments" in \cite{Carcamo2007} and subsequent researches had in fact noticed the necessary condition 2.
However, they did not state this condition explicitly like what is done here. \\

In SM, the Lagrangian as a whole is assumed to be Hermitian.
If we assume the Yukawa couplings or equivalently the $M$ matrices were also Hermitian, their patterns will be simplified remarkably.
As shown in one of our previous investigations \cite{Lin2019}, such an assumption can reduce those eighteen parameters in Eq.(4) down to only five.  \\

In the first stage, the Hermitian condition $M=M^{\dagger}$ acquires that $D_j =0$, $B_{j+3}=B_j$ and $C_{j+3}=-C_j$, for $j$ = 1, 2 and 3.
Thus, Eq.(4) becomes
\begin{eqnarray}
M &=&  \left( \begin{array}{ccc} A_1  & B_1+ i C_1 & B_2+ i C_2   \\
       B_1 - i C_1 & A_2  & B_3 + i C_3  \\ B_2 - i C_2 & B_3 - i C_3 & A_3  \end{array}\right) \nonumber \\
 &=& M_R + M_I =  \left( \begin{array}{ccc} A_1 &  B_1 &  B_2  \\  B_1 &  A_2 &  B_3  \\ B_2 & B_3 & A_3 \end{array}\right)+
      \left( \begin{array}{ccc} 0 & i C_1 & i C_2   \\  -i C_1 & 0 & i C_3  \\ -i C_2 & -i C_3 & 0 \end{array}\right),
\end{eqnarray}
where $M_R$ and $M_I$ are the real and imaginary parts of $M$, respectively.
Obviously, $M_R$ and $M_I$ are also respectively Hermitian.
Thus, the number of parameters in $M$ is reduced from eighteen down to nine. \\

If there were a unitary matrix $U$ which diagonalize $M_R$ and $M_I$ simultaneously and respectively,
it must also diagonalize the whole $M$ matrix and leads to the following equation which was originally given in \cite{Branco1985} and revised in \cite{Lin2019},
\begin{equation}
M_R  M_I^\dagger -  M_I M_R^\dagger = 0.
 \end{equation}
Such an equation provides us several extra relations among parameters and reduce the number of parameters further. \\

Substituting Eq.(5) into Eq.(6), we receive four equations
\begin{eqnarray}
B_1 C_1 &=&-B_2 C_2 =B_3 C_3, \\
(A_1-A_2) &=& ~~~(B_3 C_2+B_2 C_3)/ C_1,  \\
(A_3-A_1) &=& ~~~(B_1 C_3-B_3 C_1)/ C_2,  \\
(A_2 -A_3) &=& -(B_2 C_1+B_1 C_2)/ C_3,
\end{eqnarray}
which will reduce the parameter number from nine further down to five. \\

However, five parameters are still too many for us to diagonalize the $M$ matrix analytically.
In order to simplify the pattern of $M$ further, we will employ an extra assumption
\begin{equation}
A_1=A_2=A_3=A,
 \end{equation}
among the diagonal elements of $M$ and further summarize Eq.(7)-(10) in
\begin{equation}
B_1^2 =B_2^2 =B_3^2,~~~~~~ C_1^2 = C_2^2 =C_3^2.
\end{equation}
\\

By examining all possible solutions of Eq.(12) we found only four cases satisfy it.
In what follows they will be studied respectively as:

\subsection*{Case 1: $\bold{B_1=B_2=B_3=B}$ and $\bold{C_1=-C_2=C_3=C~~~~}$}

In this case,
\begin{eqnarray}
M = M_R + M_I= \left( \begin{array}{ccc} A   &   B  &  B  \\  B   &  A   &  B  \\  B   &   B  &  A \end{array}\right)
   + i \left( \begin{array}{ccc} 0  &   C  &  -C \\ -C   &  0   &  C  \\  C   & - C  &  0 \end{array}\right),
\end{eqnarray}
which has the same pattern as what was achieved in \cite{Lin1988, Lee1990} with a $S_3$ permutation symmetry imposed among the Yukawa interactions of three quark generations.
That symmetry was originally employed to solve the FCNC problem in a 2HDM.
However, we find it applies to SM also.    \\

Analytically, the $M$ matrix is diagonalized and the mass eigenvalues are given as
\begin{eqnarray}
M_{diag.} = \left( \begin{array}{ccc} A-B-\sqrt{3} C  & 0 &  0  \\  0  & A-B+\sqrt{3} C &  0  \\ 0  & 0 & A+2B \end{array}\right),
\end{eqnarray}
 and the corresponding $U$ matrix which diagonalize Eq.(13) is then given as
\begin{eqnarray}
U_1 = \left( \begin{array}{ccc}  {{-1-i \sqrt{3}} \over {2 \sqrt{3}}} &  {{-1+i \sqrt{3}} \over {2 \sqrt{3}}} & {1\over \sqrt{3}}  \\
{{-1+i \sqrt{3}} \over {2 \sqrt{3}}} &   {{-1-i \sqrt{3}} \over {2 \sqrt{3}}} &   {1\over \sqrt{3}}  \\
{1\over \sqrt{3}} & {1\over \sqrt{3}} & {1\over \sqrt{3}}  \end{array}\right),
\end{eqnarray}
where the sub-index $k$ ($k$ = 1 to 4) of $U_k$ indicates to which case it corresponds. \\

\subsection*{Case 2: $\bold{B_1=B_2=-B_3=B}$ and $\bold{C_1=-C_2=-C_3=C}$}

In this case,
\begin{eqnarray}
M =  M_R + M_I=  \left( \begin{array}{ccc} A   & B  &  B  \\  B   & A  &  -B  \\ B   & -B  &  A \end{array}\right)
      + i \left( \begin{array}{ccc} 0   & C  & -C   \\  -C  & 0  &  -C  \\ C  & C  &  0 \end{array}\right),
\end{eqnarray}
which reveals a residual $S_2$ symmetry between the second and third generations. \\

The mass eigenvalues are given as
\begin{eqnarray}
M_{diag.} = \left( \begin{array}{ccc} A+B-\sqrt{3} C  & 0 &  0  \\  0  & A+B+\sqrt{3} C &  0  \\ 0  & 0 & A-2B \end{array}\right),
\end{eqnarray}
and the corresponding $U$ matrix is given as
\begin{eqnarray}
U_2 = \left( \begin{array}{ccc}
  {{1-i \sqrt{3}} \over {2 \sqrt{3}}} & {{-1-i \sqrt{3}} \over {2 \sqrt{3}}} &   {1\over \sqrt{3}} \\
 {{1+i \sqrt{3}} \over {2 \sqrt{3}}} &  {{-1+i \sqrt{3}} \over {2 \sqrt{3}}} & {1\over \sqrt{3}} \\
  {-1\over \sqrt{3}} & {1\over \sqrt{3}} & ~{1\over \sqrt{3}} \end{array}\right).
\end{eqnarray}

\subsection*{Case 3: $\bold{B_1=-B_2=B_3=B}$ and $\bold{C_1=C_2=C_3=C}$}

In this case,
\begin{eqnarray}
M =  M_R + M_I= \left( \begin{array}{ccc} A   & B  &  -B  \\  B   & A  &  B  \\ -B   & B  & ~ A \end{array}\right)
       +  i \left( \begin{array}{ccc} 0   & C  &  C   \\  -C  & 0  & C  \\ -C   & -C  & 0 \end{array}\right),
\end{eqnarray}
which reveals a residual $S_2$ symmetry between the first and third generations. \\

The mass eigenvalues are given as
\begin{eqnarray}
M_{diag.} = \left( \begin{array}{ccc} A+B-\sqrt{3} C  & 0 &  0  \\  0  & A+B+\sqrt{3} C &  0  \\ 0  & 0 & A-2B \end{array}\right),
\end{eqnarray}
and the corresponding $U$ matrix is given as
\begin{eqnarray}
U_3 = \left( \begin{array}{ccc}   {{-1+i \sqrt{3}} \over {2 \sqrt{3}}} & {{1+i \sqrt{3}} \over {2 \sqrt{3}}} &  {1\over \sqrt{3}} \\
{{-1-i \sqrt{3}} \over {2 \sqrt{3}}} &  {{1-i \sqrt{3}} \over {2 \sqrt{3}}} & {1\over \sqrt{3}} \\
  {1\over \sqrt{3}} & {-1\over \sqrt{3}} & {1\over \sqrt{3}}   \end{array}\right).
\end{eqnarray}

\subsection*{Case 4: $\bold{B_1=-B_2=-B_3=-B}$ and $\bold{C_1=C_2=-C_3=-C}$}

In this case,
\begin{eqnarray}
M = M_R + M_I= \left( \begin{array}{ccc} A   & -B  &  B  \\  -B   & A  &  B  \\ B   & B  &  A \end{array}\right)
 + i \left( \begin{array}{ccc} 0   & -C  &  -C   \\  C   & 0  & C  \\ C   & -C  & 0 \end{array}\right),
\end{eqnarray}
which reveals a residual $S_2$ symmetry between the first and second generations. \\

The mass eigenvalues are given as
\begin{eqnarray}
M_{diag.} = \left( \begin{array}{ccc} A+B-\sqrt{3} C  & 0 &  0  \\  0  & A+B+\sqrt{3} C &  0  \\ 0  & 0 & A-2B \end{array}\right),
\end{eqnarray}
and the corresponding $U$ matrix is then given as
\begin{eqnarray}
U_4 = \left( \begin{array}{ccc}  {{1-i \sqrt{3}} \over {2 \sqrt{3}}} & {{1+i \sqrt{3}} \over {2 \sqrt{3}}} & {1\over \sqrt{3}} \\
{{1+i \sqrt{3}} \over {2 \sqrt{3}}} & {{1-i \sqrt{3}} \over {2 \sqrt{3}}} & {1\over \sqrt{3}} \\
 {1\over \sqrt{3}} & {-1\over \sqrt{3}} & {1\over \sqrt{3}} \end{array}\right).
\end{eqnarray}
\\

In all four cases the $U_k$ matrices are complex which satisfy the first condition stated in section I. If we assign different of them to up- and down-type quarks respectively, the second condition will also be satisfied. Even so, we are still not sure if $V_{CKM}$ were complex since the inner product of two complex matrices can still be real, which is to be demonstrated in next section. \\

In fact, derivations presented above were originally proposed to solve the FCNC problem in the 2HDM by finding matrix pairs which can be diagonalized by a same $U$ matrix simultaneously \cite{Lin2013}.
However, we found in some special cases the derived CKM matrices are complex which mean that CP symmetry is violated. Besides, the derivations apply not only to 2HDMs, but to the SM also. As SM is already enough to give CPV explicitly, why should we bother ourselves to deal with the FCNC problem in the 2HDM? Surely we can still apply them to the extensions of SM with one or even two extra Higgs doublets while the FCNC problem vanishes naturally at tree level. But that is a different subject to be discussed elsewhere.  \\

Besides, it is noteworthy those $S_N$ symmetries are not imposed constraints.
They are by-products brought in by the assumption among $A$ parameters in Eq.(11).
Among them, the $S_3$-symmetric pattern in ${\bf case~1}$ is exactly the same as the one derived in our very early manuscripts \cite{Lin1988, Lee1990}.
It solved the FCNC problem at tree level successfully but not the CPV problem.
The problem it met at that time was the breach of ${\bf condition~ 2}$, $U^u \neq U^d$, since we had only one $S_3$-symmetric $U_1$ matrix for both quark types.
While in this manuscript, extra $S_2$-symmetric $U_2$, $U_3$ and $U_4$ matrices give us a key to ignite violation of CP symmetry from the theoretical end.  \\

\section{The CKM matrix and the BAU problem}

As mentioned in section I, if one expects to yield a CP-violating phase in $V_{CKM}$, two necessary conditions are to be satisfied.
In last section, four complex $U_k$ matrices were achieved with a Hermitian assumption of $M$ and an assumption among $A$ parameters.
If we assign different $U_k$ matrices to up- and down-type quarks respectively, both conditions are satisfied.
In what follows, various assignments of $U_k^u$ and $U_k^d$ are examined and $V_{CKM}$ they yield are presented in TABLE. I.
Several of them are complex which indicate a theoretical origin of CPV is explicitly yielded. \\

The full expressions of matrices ${\bf{1_{3\times3}}}$, ${\bf E}$, ${\bf F}$, ${\bf G}$ and ${\bf H}$ in TABLE I are given as what follows
\begin{eqnarray}
{\bf 1_{3\times 3}} &=& \left( \begin{array}{ccc} 1 & 0 & 0  \\  0 & 1 & 0 \\ 0 & 0 & 1 \end{array}\right),
~\bold{G} = \left( \begin{array}{ccc} 2\over 3 & -1\over 3 & 2\over 3  \\  -1\over 3 & 2\over 3 & 2\over 3  \\ 2\over 3 & 2\over 3 & -1\over 3 \end{array}\right),
~\bold{H} = \left( \begin{array}{ccc} -1\over 3 & 2\over 3 & 2\over 3   \\ 2\over 3 & -1\over 3 &  2\over 3  \\ 2\over 3 & 2\over 3 & -1\over 3 \end{array}\right), \nonumber \\
\bold{E} &=& \left( \begin{array}{ccc} {{1-i \sqrt{3}}\over 3} & {1\over 3}                & {{1+i \sqrt{3}}\over 3}   \\
                                                   {1\over 3}  &  {{1+i \sqrt{3}}\over 3} & {{1-i \sqrt{3}}\over 3} \\
                                    {{1+i \sqrt{3}}\over 3}    &  {{1-i \sqrt{3}}\over 3} &  {1\over 3} \end{array}\right) =
                                    \left( \begin{array}{ccc} {2 \over 3}e^{-i {\pi \over 6}}  & {1 \over 3} & {2 \over 3}e^{i {\pi \over 6}} \\
            { 1\over 3} & {2 \over 3}e^{i {\pi \over 6}} & {2 \over 3}e^{-i {\pi \over 6}} \\
                  {2 \over 3}e^{i {\pi \over 6}} & {2 \over 3} e^{-i {\pi \over 6}} & {1 \over 3}\end{array}\right), \nonumber \\
\bold{F} &=& \left( \begin{array}{ccc} {1\over 3} &  {{1-i \sqrt{3}}\over 3} & {{1+i \sqrt{3}}\over 3}   \\
                          {{1+i \sqrt{3}}\over 3} &  {1\over 3} &  {{1-i \sqrt{3}}\over 3}  \\
                               {{1-i \sqrt{3}}\over 3} &  {{1+i \sqrt{3}}\over 3} & 1 \over 3   \end{array}\right)
                               =\left( \begin{array}{ccc} {1 \over 3} & {2 \over 3} e^{-i {\pi \over 6}} & {2 \over 3}e^{i {\pi \over 6}} \\
             {2 \over 3}e^{i {\pi \over 6}} & { 1\over 3} & {2 \over 3}e^{-i {\pi \over 6}} \\
              {2 \over 3}e^{-i {\pi \over 6}} & {2 \over 3} e^{i {\pi \over 6}} & {1 \over 3}\end{array}\right).
\end{eqnarray}
The matrices $\bold{1_{3\times 3}}$, $\bold{G}$ and $\bold{H}$ are purely real and obviously CP-conserving.
While $\bold{E}$,  $\bold{F}$ and their complex conjugates are complex, which means they are CP-violating.
It is interesting that as shown in the $\bold{G}$ and $\bold{H}$ cases, even if both conditions were satisfied, $V_{CKM}$ can still be completely real.  \\

\begin{table}[tbp]
\centering
\begin{tabular}{|l|llll|}
\hline
$V_{CKM}$ & ~~$U_1^{d \dagger}$ & $U_2^{d \dagger}$ & $U_3^{d \dagger}$ & $U_4^{d \dagger}$ \\
\hline
~~$U^u_1$ & ~~ $1_{3\times 3}$    &   $E$                & $E^*$           &      $G$    \\
~~$U^u_2$ & ~~ $E^*$              &   $1_{3\times 3}$    & $H$             &      $F$  \\
~~$U^u_3$ & ~~ $E$                &   $H$                & $1_{3\times 3}$ &      $F^*$    \\
~~$U^u_4$ & ~~ $G$                &   $F$                & $F^*$           &      $1_{3\times 3}$ \\
\hline
\end{tabular}
\caption{\label{tab:i} Various assembles of CKM matrix.}
\end{table}

We now have already a way to describe the violation of CP symmetry from the theoretical end. But the CKM elements derived in Eq.(25) are very different to experimentally detected corresponding values.
Some of them are hundreds times the detected values, say both predicted $\vert V_{ub} \vert = 2/3$ in ${\bf E}$ and ${\bf F}$ are about 184 times the presently detected value $3.82 \pm 0.24 \times 10^{-3}$ \cite{Zyla2020}.
Besides, the CKM matrices derived here contain only numbers rather than any variable factors.
That leaves us no space to improve the fitting between theoretical predictions and experimental detections further.
This could be ascribed to the over-simplified matrix patterns caused by the $A=A_1=A_2=A_3$ assumption in Eq.(11) or equivalently the revealed $S_N$ symmetries.
However, it may also hint that if we can diagonalize Eq.(5) directly rather than imposing the assumption in Eq.(11),
we will achieve $M$ patterns more close to the ones observed now. \\

As we have already yield several complex $V_{CKM}$ demonstrated in Table I, it is rational for us to go one step further to estimate the CPV strength predicted by such a model.
In usual, the strength of direct CPV is estimated by the dimensionless Jarlskog determinant  \cite{Jarlskog1985, Farrar1993, Farrar1994, Tranberg2010} which was given as
\begin{eqnarray}
\Delta_{CP} &=& v^{-12} {\rm Im~ Det} [m_u m_u^{\dagger} , m_d m_d^{\dagger} ] \nonumber \\
            &=& {\it J}~ v^{-12} \prod_{\scriptstyle i < j} (\tilde{m}_{u,i}^2 - \tilde{m}_{u,j}^2 ) \prod_{\scriptstyle i < j}(\tilde{m}_{d,i}^2 - \tilde{m}_{d,j}^2 ) \simeq 10^{-19},
\end{eqnarray}
where $v$ is the Higgs vacuum expectation value and $\tilde{m}$ are particle masses. \\

The Jarlskog invariant ${\it "J"}$ in Eq.(26) is a phase-convention-independent measure of CPV defined by
\begin{eqnarray}
{\rm Im} [V_{ij} V_{kl} V_{il}^* V_{kj}^* ]={\it J}~ \Sigma_{m,n} \epsilon_{ikm} \epsilon_{jln},
\end{eqnarray}
and a global fit of its value was given by the ${\bf Particle~ Data~ Group}$ as ${\it J}=(3.00^{+0.15}_{-0.09}) \times 10^{-5}$ \cite{Zyla2020}.
Such a value corresponds to a BAU of the order $\eta \sim 10^{-20}$ if one substitutes the detected fermion masses into Eq.(26).
It is obviously too small to account for the cosmologically observed $\eta ={N_B \over N_{\gamma}} \vert_{T=3K} \sim 10^{-10}$,
where $N_B$ is the number of baryons and $N_{\gamma}$ is the number of photons.
 \\

Substituting the elements $V_{cd}$, $V_{ts}$, $V_{cb}^*$ and $V_{us}^*$ given in Eq.(25) into (27),
the ${\it J}$ values for ${\bf F}$ and ${\bf F^*}$ are the same
\begin{eqnarray}
 {\it J}_{\bf F,~F^*}={16 \over 81} \sin(2\pi /3) \sim 0.171,
\end{eqnarray}
while those for ${\bf E}$ and ${\bf E^*}$ are zero since the phases in them are canceled.
The ${\it J}$ value given in Eq.(28) is almost four orders higher than the one given by current SM.
It hints that under circumstances with $S^{(c \leftrightarrow t)}_2+S^{(d \leftrightarrow s)}_2$, $S^{(u \leftrightarrow t)}_2+S^{(d \leftrightarrow s)}_2$, $S^{(u \leftrightarrow c)}_2 +S^{(d \leftrightarrow b)}_2$ and $S^{(u \leftrightarrow c)}_2+S^{(s \leftrightarrow b)}_2$ symmetries,
the hyper-indices in the brackets indicate between which two quarks the $S_2$ symmetry appears,
the CPV strengths thus derived are orders stronger than what we see nowadays and thus more productive of BAU. \\

But, as constrained by Eq.(27), the value of $J$ is always smaller than 1.
So a large $J$ is always not enough to account for all the discrepancy between the cosmologically observed BAU and the one current SM predicts.
One of the other potential sources of a large $\Delta_{CP}$ is the mass-square-depending terms $(\tilde{m}_{u,i}^2 - \tilde{m}_{u,j}^2 )$ and $(\tilde{m}_{d,i}^2 - \tilde{m}_{d,j}^2 )$ in Eq.(26).
Unfortunately, the mass eigenvalues derived here give no helps on this subject since they are fixed to the detected quark masses.
Obviously, finding mass eigenvalues running with some variables in such models will be very helpful in solving the BAU problem. \\

Besides, symmetries usually exist under circumstances with higher temperatures.
As we do not see such $S_N$ symmetries among fermion generations in our present universe,
it is natural for us to consider they could have appeared in some early eras if the Big-Bang cosmology were correct.
Thus, the discrepancy between the $S_N$-symmetric CPV and that detected in present experiments is also natural. \\

In another way, we may imagine in some very early stages of the universe with extremely high temperature ${\it T}$ there were $S_3$ symmetries in all fermion types.
As ${\it T}$ decreased with the expansion, some fermions degenerated from others and the symmetries were broken down to a residual $S_2$-one.
Probably the up-type quarks first and then the down-type quarks followed, but this is not necessary.
During a succession of breaking down of $S_N$ symmetries, for instance
\begin{eqnarray}
S_3^u + S_3^d &\rightarrow& ~S_2^u~+S_3^d ~\rightarrow ~S_2^u ~+~S_2^d \\ \nonumber
              &\rightarrow& ~No^u ~+S_2^d ~\rightarrow No^u ~+No^d ~,
\end{eqnarray}
where the super-indices $u$ and $d$ indicate up- and down-type quarks respectively.
At least the $S_2^u~+S_2^d$ era was proved to be CP-violating and very productive of BAU in this manuscript. \\

s we do not see any $S_N$ symmetries in our present universe, obviously we are now in the latest stage, ${\it No^u +No^d}$, of Eq.(29).
That means a completely analytical diagonalization of $M$ matrices is desired.
We expect such a diagonalization will give us a $V_{CKM}$ coincides the experimentally detected values better.
Unfortunately, this is still unaccomplished for now and will be the next goal of our future investigations. \\

\section{Conclusions and Discussions}

Since the discovery of CPV in decays of neutral kaons, its theoretical origin is always a puzzle physicists urgent to solve.
Through the analysis on compositions of $V_{CKM}$ two necessary but not sufficient conditions for yielding a complex $V_{CKM}$ are stated and an explicit way to describe the violation of CP symmetry in SM is presented. \\

The way performed in this manuscript is not a new one to study the CP problem.
It's in fact an old and orthodox way in SM,
by diagonalizing the $M$ matrices to obtain corresponding $U$ matrices and then yields a complex $V_{CKM}$.
In previous attempts we failed in this goal since most researchers ignored the second condition while in this manuscript that condition is satisfied since four such $U_k$ matrices are achieved.

The key factor enables us to achieve a manageable $M$ pattern and consequently a CP-violating $V_{CKM}$ is Eq.(6).
Such a condition between the real and imaginary parts of a Hermitian matrix is always true as proved in \cite{Lin2019} if they were diagonalized by a same $U$ matrix.
It correlates the elements in an $M$ matrix and thus reduces the number of parameters from eighteen down to five.
Furthermore, with an extra assumption among its parameters as given in Eq.(11) four $S_N$-symmetric $M_k$ patterns are achieved in section II.
Thus, the second condition is also satisfied and several complex CKM patterns come in explicitly. \\

Though the magnitudes of Jarlskog invariants thus derived are about four orders that predicted by current SM.
However, as we do not see any such $S_N$ symmetries in our present universe and they should exist only under circumstances with extremely high temperature.
It's natural to consider their appearances in some very early stages of the universe.
That hints our universe could have been very productive of BAU in such $S_N$-symmetric stages.
While, as the $S_N$ symmetries broke down completely as the universe expands,
such a BAU-productive mechanism was narrowed down to the present status, ${\it J}=(3.00^{+0.15}_{-0.09}) \times 10^{-5}$.
Obviously, such an improved SM not only describes how CP symmetry was violated from the theoretical end,
but also explains why the amount of BAU observed today is so huge while the currently detected CPV is so tiny  \\

This work is not a new model. Instead, it is just an improvement of SM in the sector of fermions' Yukawa interactions.
Though it does not solve the CPV and BAU problems completely, it still improves our understanding of the origin of CPV and exhibits a way to account for partly the cosmologically observed BAU which cannot be accounted for by current SM.
Such an interesting aspect could be a good stepping stone for coming researchers on these topics. \\

\end{document}